\documentclass[11pt,a4paper]{article}%
\usepackage[T1]{fontenc}
\usepackage[latin1,utf8]{inputenc}

\pdfoutput=1

\usepackage{amsmath,amssymb,amsfonts,multicol}
\usepackage[normal,font=small,labelfont=bf,labelsep=period]{caption}
\usepackage[pdftex]{color,graphicx}
\usepackage[english]{babel}
\usepackage[compress]{cite}
\usepackage{color}
\usepackage{graphicx}
\usepackage{cancel}
\usepackage{amsfonts}
\usepackage{amssymb}
\usepackage{amsmath}
\usepackage{calc}
\usepackage{dcolumn}
\usepackage{mathrsfs}
\usepackage{mathtools}
\usepackage{soul}
\usepackage{mathtools}
\usepackage{braket}
\usepackage[normalem]{ulem}
\usepackage{url}
\usepackage[dvipsnames]{xcolor}
\definecolor{darkgreen}{rgb}{0,0.65,0}
\usepackage{slashed}

\usepackage[linktoc=all,colorlinks=true,linkcolor=black,urlcolor=magenta,citecolor=black]{hyperref}
\hypersetup{%
  colorlinks = true,
  linkcolor  = black
}
\definecolor{rossoCP3}{cmyk}{0,.88,.77,.40}
\definecolor{verdeCP3}{rgb}{0.09765625, 0.57421875, 0.1015625}
\definecolor{bluCP3}{rgb}{0, 0.23, 0.67}
\ifx\pdfoutput\undefined
\usepackage[dvips,bookmarks]{hyperref}    
\else
\usepackage{hyperref}    
\fi
\hypersetup{colorlinks, bookmarksopen, bookmarksnumbered,
citecolor=verdeCP3, linkcolor=bluCP3, pdfstartview=FitH, urlcolor=rossoCP3}

\newcommand{\be}{\begin{equation}}
\newcommand{\ee}{\end{equation}}
\newcommand{\bea}{\begin{eqnarray}}
\newcommand{\eea}{\end{eqnarray}}

\newcommand{\ie}{i.e.~}
\newcommand{\eg}{e.g.~}

\newcommand{\Eq}[1]{Eq.~\eqref{#1}}
\newcommand{\Fig}[1]{Fig.~\ref{#1}}
\newcommand{\Tab}[1]{Tab.~\ref{#1}}
\newcommand{\Sec}[1]{Sec.~\ref{#1}}

\definecolor{corr}{rgb}{1,0,.2}

\definecolor{gb}{rgb}{0.8,0.3,0.1}

\numberwithin{equation}{section}

\begin{document}

\color{black}

\begin{center}
{\LARGE\color{black}\bf Detecting the Stimulated Decay of Axions at Radio Frequencies\\[1mm] }

\medskip
\bigskip\color{black}\vspace{0.6cm}

{
{\large\bf Andrea Caputo$^{a}$,}\ 
{\large\bf Marco Regis$^{b,c}$,}\ 
{\large\bf Marco Taoso$^{c}$,}\ 
{\large\bf Samuel J. Witte$^{a}$}\ 
}
\\[7mm]

{\it $^a$  Instituto de F\'{i}sica Corpuscular, CSIC-Universitat de Valencia,
Apartado de Correos 22085, E-46071, Spain
}\\[3mm]
{\it $^b$  Dipartimento di Fisica, Universit\`{a} di Torino, via P. Giuria 1, I--10125 Torino, Italy}\\[3mm]
{\it $^c$  Istituto Nazionale di Fisica Nucleare, Sezione di Torino, via P. Giuria 1, I--10125 Torino, Italy}\\[3mm]
{E-mail: andrea.caputo@uv.es,}
{regis@to.infn.it,}
{marco.taoso@gmail.com,}
{witte.sam@gmail.com}
\end{center}

\bigskip

\centerline{\large\bf Abstract}
\begin{quote}
\color{black}\large 
Assuming axion-like particles account for the entirety of the dark matter in the Universe, we study the possibility of detecting their decay into photons at radio frequencies. 
We discuss different astrophysical targets, such as dwarf spheroidal galaxies, the Galactic Center and halo, and galaxy clusters. 
The presence of an ambient radiation field leads to a stimulated enhancement of the decay rate; depending on the environment and the mass of the axion, the effect of stimulated emission may amplify the photon flux by serval orders of magnitude. 
For axion-photon couplings allowed by astrophysical and laboratory constraints (and possibly favored by stellar cooling),
we find the signal to be within the reach of next-generation radio telescopes such as the Square Kilometer Array. 

\end{quote}

\newpage

\section{Introduction}
\label{sec:introduction}
 
The cumulative astrophysical and cosmological evidence for the existence of a non-baryonic, minimally interacting, cold matter component of the Universe (conventionally referred to as dark matter) is 
overwhelming, with current observations suggesting that dark matter resides in the form of new unknown particles.
However, the exact nature of dark matter continues to evade physicists. 

The most popular dark matter candidates are those naturally capable resolving additional fundamental problems at the forefront of particle physics. One such candidate is the QCD axion, which inherently appears in the Peccei-Quinn solution to the strong CP problem\cite{Peccei:1977hh,Peccei:1977ur, Weinberg:1977ma, Wilczek:1977pj}\footnote{Simply put, the strong CP problem arises from the fact that the QCD $\theta$ term predicts a non-vanishing neutron electric dipole moment, while current experimental bounds constrain $\theta$ to an unnaturally small value, $\theta \lesssim 10^{-10}$. }. 
If the mass of the axion is $\lesssim 20$ eV, the axion is stable on cosmological timescales and can contribute substantially to 
the current fraction of energy density in the Universe stored in form of cold dark matter~\cite{Preskill:1982cy, Dine:1982ah, Abbott:1982af, Davis:1986xc}.
Astrophysical observations constrain the axion mass to reside approximately between $10^{-4}\,\mu{\rm eV}$ and $10^4\,\mu{\rm eV}$~\cite{Marsh:2017hbv,Irastorza:2018dyq}~\footnote{However see~\cite{DiLuzio:2017ogq} for models where the astrophysical bounds are relaxed.}. 
The mass range where axions can account for the entirety of the dark matter depends on the interplay between different production mechanisms (in particular, the interplay between the misalignment mechanism and the decay of topological defects) and
whether the PQ symmetry (\ie the symmetry introduced in the Peccei-Quinn solution to the strong CP problem) is broken before or after inflation (see \eg~\cite{Marsh:2017hbv,Irastorza:2018dyq} for more extensive discussions). In the post-inflationary PQ breaking scenario, and assuming axions are produced exclusively from the misalignment mechanism, one finds the axion mass $m_a \simeq \rm{few }\times(10)\,\mu{\rm eV}$~\cite{Borsanyi:2016ksw}.
As we will show, radio telescopes searching for axion decay are ideally placed to probe this mass regime.

In recent years an increasing amount of attention has shifted toward searching for axion dark matter. 
An important consequence of this has been the development of a diverse and complementary search program intended to probe the many unique facets of such a dark matter candidate; this program includes, but is not limited to, haloscopes \cite{Hagmann:1990tj, Hagmann:1998cb, Asztalos:2001tf, Du:2018uak, Zhong:2018rsr}, heliscopes\cite{Anastassopoulos:2017ftl,Irastorza:2013dav,Armengaud:2014gea}, $5^{\rm th}$ force experiments~\cite{Vasilakis:2008yn,Raffelt:2012sp,Heckel:2013ina,Terrano:2015sna,Geraci:2017bmq}, light-shinning-through wall experiments~\cite{Cameron:1993mr,Robilliard:2007bq,Chou:2007zzc,Afanasev:2008jt,Ehret:2010mh,Ballou:2015cka}, LC circuit resonators~\cite{Sikivie:2013laa,Kahn:2016aff,Silva-Feaver:2016qhh}, oscillating nuclear dipole searches~\cite{Graham:2011qk,Budker:2013hfa,Barbieri:2016vwg}, axion-induced atomic transitions~\cite{Sikivie:2014lha, Flambaum:2018ssk}, axion-induced atomic and molecular electric dipole moments \cite{Stadnik:2017hpa}, and indirect axion searches~\cite{Kelley:2017vaa,Sigl:2017sew,Huang:2018lxq,Hook:2018iia,Safdi:2018oeu,Caputo:2018ljp}. 
Many of these searches, although certainly not all, rely on the axion's coupling to photons; this interaction is given by the operator $\mathcal{L}=-\frac{1}{4}g_{a\gamma\gamma}\,a\,F_{\mu\nu}\tilde{F}_{\mu\nu},$
where $a$ is the axion field, $F_{\mu\nu}$ is the electromagnetic field strength, $\tilde{F}_{\mu\nu}$ its dual, and $g_{a\gamma\gamma}$ the coupling constant. Of particular importance here is the notion that one may be able to exploit the large number density of axions in astrophysical environments to indirectly infer their existence through the detection of low-energy photons.  For axion masses in the `characteristic' dark matter window (\ie $\mu{\rm eV} \lesssim m_a \lesssim  10^2 \mu{\rm eV}$), the energy of a non-relativistic axion corresponds to a photon with a frequency ranging from $\sim \mathcal{O}(100)$ MHz to $\sim \mathcal{O}(10)$ GHz; intriguingly, this lies exactly in the range of frequencies probed by radio telescopes.

There have been various attempts in recent years to use radio telescopes to detect axion dark matter, a majority of which have relied on the axion-to-photon conversion process (\ie the so-called Primakoff effect). Recently it was shown that unless one exploits a resonant axion-photon conversion (as \eg was done in~\cite{Hook:2018iia,Huang:2018lxq,Safdi:2018oeu}), the rate of axion decay into two photons will likely supersede that of axion-photon conversion in large-scale astrophysical environments\cite{Sigl:2017sew,Caputo:2018ljp}. One of the difficulties with resonant searches is that they rely on a comprehensive understanding of highly uncertain astrophysical environments. An alternative approach with a far more limited dependence on astrophysical uncertainties, albeit at the potential cost of sensitivity, was proposed in~\cite{Caputo:2018ljp}. Ref~\cite{Caputo:2018ljp} performed an exploratory study for an idealized near future radio telescope to determine whether the axion-to-two-photon decay process could potentially produce an observable signature. This work presented here is intended to serve as a comprehensive follow-up, incorporating a far more sophisticated treatment of near-future radio sensitivity and exploring a variety of astrophysical sources (including the Galactic halo, the galaxy M87 in the Virgo cluster and dwarf spheroidal galaxies). 
Moreover, this work presents a more detailed treatment of the stimulated emission, a mechanism which is induced by the presence of a background radiation in the medium where the axion decay occurs (see also \cite{Rosa:2017ury, Tkachev:1987cd, Kephart:1994uy}).
At low radio frequencies, this effect enhances the expected emission by several orders of magnitude.

The coupling of the QCD axion to the photon $g_{\gamma\gamma}$ grows linearly with the axion mass and with a proportionality constant that depends on the UV completion of the axion model. Considering various types of UV completions thus defines a band in the mass-coupling plane identifying where viable QCD axions may reside (see the light green region in~\Fig{fig:resultsDW})\footnote{See~\cite{DiLuzio:2016sbl} for a recent determination of the allowed region in the $g_{\gamma\gamma}-m_a$ plane for the KSVZ model\cite{Zhitnitsky:1980tq, Dine:1981rt} and~\cite{Agrawal:2017cmd,Farina:2016tgd} for scenarios where the  range of couplings can be further extended.}.
More generically, many extensions of the Standard Model predict light particles with similar properties to the QCD axion, but that might not be related to the strong CP problem and for which the relation between $g_{\gamma\gamma}$ and the mass could be different.
These are referred to as axion-like particles (ALPs), and they appear generically in low energy-effective theories arising from string theory, e.g.~\cite{Witten:1984dg,Svrcek:2006yi,Arvanitaki:2009fg,Cicoli:2013ana}.
It is therefore important to explore all the parameter space of ALPs, beyond the well-motivated case of the QCD axion.

The results presented here suggest that near-future surveys will be incapable of probing the parameter space of the QCD axion; however we find that the Square Kilometer Array (SKA) will be able to improve current bounds on ALPs by about one order of magnitude.

The paper is organized as follows. Section~\ref{sec:stimulated} outlines the origin of stimulated decay and the relevant contributions to the ambient photon background.  
Section~\ref{sec:senstivities} describes how to compute the expected signal-to-noise arising from axion decay for a generic choice of astrophysical environment and radio telescope. We present the sensitivity
on the ALP-photon coupling for various astrophysical targets and telescopes in Section~\ref{sec:results}. In Section~\ref{sec:conclusions} we conclude.

\section{Axion decay in a photon bath - stimulated emission}
\label{sec:stimulated}
The decay of an axion with mass $m_a$ proceeds through the chiral anomaly and produces two photons, each with a frequency $\nu=m_a/4\pi$. The lifetime of the axion can be expressed in terms of its mass and the effective axion-two-photon coupling $g_{a\gamma\gamma}$ as
\begin{equation}
	\tau_a=\frac{64\pi}{m_a^3 \,g_{a\gamma\gamma}^2} \, .
\label{eq:tau}
\end{equation}
Evaluating the lifetime for an axion mass $m_a \sim 1\,\mu{\rm eV}$ and a coupling near the current upper limit, \ie $g_{a\gamma\gamma} \sim 10^{-10}\,{\rm GeV}^{-1}$, one finds $\tau_a \sim 10^{32}$ years; this is perhaps the main reason why axion decay has been largely neglected in the literature. The decay rate, however, is only valid in vacuum. In reality, for the axion masses of interest this decay process takes place in an ambient radiation field, which, at radio frequencies is sourced by the combination of cosmic microwave background (CMB) radiation, synchrotron radiation, and bremsstrahlung radiation. Consequently, the photon production rate is enhanced via stimulated emission, a phenomenon due to the indistinguishability of photons and Bose-Einstein statistics.
Here we review how to derive the effect in the case of a decay into two photons (for the more canonical decay into a single photon, see, e.g.~\cite{Schwartz:2013pla}).

Let us denote a particular phase-space distribution with $f$, related to the number density of particles $n$ by $dn = \frac{g}{(2\pi)^3} \,f(\bold{p})\, d^3p,$ with $g$ being the number of degrees of freedom. 
We consider a initial quantum state where there exist $f_a$ axions with a particular momentum. The final state contains $f_a-1$ axions and two photons, each with half the energy of the initial axion. The decay occurs in a medium of photons, with $f_\gamma$ particles with the same momentum and polarization as the photons produced by the axion decay.
Therefore, the initial state is $|f_a;f_\gamma;f_\gamma\rangle$ and the final state is $|f_a-1;f_\gamma+1;f_\gamma+1\rangle$.
The interaction Hamiltonian in terms of the creation and annihilation operators looks like 
\begin{equation}
H_{int}=\mathcal{M}_0^\dagger a^\dagger_\gamma a^\dagger_\gamma a_a + h.c.
\end{equation}
 where $\mathcal{M}_0$ is related to the spontaneous emission, as will be made clear momentarily. The matrix element associated with the probability of having a transition from initial to final state is:
\be
\mathcal{M}_{i\rightarrow f}=\langle f_a-1;f_\gamma+1;f_\gamma+1|H| f_a;f_\gamma;f_\gamma\rangle=\mathcal{M}_0^\dagger \sqrt{f_a}\sqrt{f_\gamma+1}\sqrt{f_\gamma+1}. 
\ee
where we have used the properties of ladder operators (i.e., $a |f_i\rangle=\sqrt{f_i}|f_i-1\rangle$ and $a^\dagger |f_i\rangle=\sqrt{f_i+1}|f_i+1\rangle$). Squaring the matrix element, one finds
\be
|\mathcal{M}_{i\rightarrow f}|^2=|\mathcal{M}_0|^2\,f_a\,(f_\gamma+1)^2 \, .
\label{eq:st1}
\ee
A similar computation can be performed for the inverse process (\ie two photons creating an axion in a medium with $f_\gamma$ photons and $f_a$ axions):
\be
\mathcal{M}_{f\rightarrow i}=\langle f_a+1;f_\gamma-1;f_\gamma-1|H|f_a;f_\gamma;f_\gamma\rangle \rightarrow |\mathcal{M}_{f\rightarrow i}|^2=|\mathcal{M}_0|^2\,f_\gamma^2\,(f_a+1)\;.
\label{eq:st2}
\ee
The variation of the number of axions is just the difference between production (described by \Eq{eq:st2}, which is the term usually identified as the ``absorption'' term) and decay of axions (given by Eq.~\ref{eq:st1}, i.e., the emission term):
\bea
|\mathcal{M}_{f\rightarrow i}|^2-|\mathcal{M}_{i\rightarrow f}|^2=-|\mathcal{M}_0|^2\,(f_a+2 f_a\,f_\gamma-f_\gamma^2)
\label{eq:st3}
\eea
The last three terms in the right-hand side of \Eq{eq:st3} describe the spontaneous decay, stimulated decay, and inverse decay, respectively.
Note that the term $\propto f_a\,f_\gamma^2$ cancels out between decay and production.
For all environments considered in this work $f_a\gg f_\gamma$, so the inverse decay can be neglected. 

The axion decay rate can then be obtained integrating the matrix element over all the momenta (and imposing energy-momentum conservation), leading to the well-known Boltzmann Equation (see e.g.~\cite{Kolb:1990vq}):
\bea \label{eq:boltz}
	\dot{n}_a&=& -\int d\Pi_a d\Pi_\gamma d\Pi_\gamma (2\pi)^4\,\delta^4(p_a-p_\gamma-p_\gamma)\, |\mathcal{M}_0|^2 \,[f_a\,(f_{\gamma}+1)(f_{\gamma}+1)+(f_a+1)\,f_{\gamma}\,f_{\gamma}]\nonumber\\
                        &=& -\int d\Pi_a d\Pi_\gamma d\Pi_\gamma (2\pi)^4\,\delta^4(p_a-p_\gamma-p_\gamma)\, |\mathcal{M}_0|^2 \,[f_a\,(1+2\,f_{\gamma})-f_{\gamma}^2]\nonumber\\
                        &\simeq&-n_a\,\Gamma_a\,(1+2\,f_{\gamma})\;,
\eea
where $d\Pi_i=g_i/(2\pi)^3 d^3p/(2\,E)$ 
and in the last line we used the definition of the decay rate $\Gamma_a=\int d\Pi_\gamma d\Pi_\gamma (2\pi)^4\,\delta^4(p_a-p_\gamma-p_\gamma)\, |\mathcal{M}_0|^2/(2\,m_a)$. 

It should be clear from \Eq{eq:boltz} that the effect of stimulated emission can be incorporated by simply multiplying the rate of spontaneous emission by a factor $2\,f_{\gamma}$. 
The photon occupation number $f_{\gamma}$ can be obtained from the associated differential energy density of the ambient radiation using $\rho_i(E_i)\,dE_i=d\Pi_i\,2\,E_i^2\, f_i $ which leads to: 
\be 
f_\gamma=\frac{\pi^2\rho_\gamma}{E_\gamma^3}.
\label{eq:ffromI}
\ee

In Fig.~\ref{fig:stimulated}, we show the stimulated emission factor arising from the CMB (black), Galactic diffuse emission (red), and the extragalactic radio background (green), as a function of the axion mass. Fig.~\ref{fig:stimulated} shows that for an axion mass $m_a \sim 1\,\mu{\rm eV}$, the stimulated decay produces an enhancement by a factor $\gtrsim 10^5$, regardless of the astrophysical environment.

\section{Radio Sensitivity}
\label{sec:senstivities}

In the following Sections we outline the procedure for computing the expected radio emission from, and the detectability of, axion decay for various astrophysical targets and telescopes. Before beginning, we comment on a number of subtle, but important features of the expected signal. First, the signal is expected to be rather diffuse, at least when compared with the beam of typical radio telescopes. For such a diffuse emission, a relatively large beam is desired in order to enhance the signal-to-noise. This can be easily achieved by single-dish telescopes. On the other hand, their collecting area (and thus the sensitivity) is typically much smaller than that of interferometers (that, in addition, have a smaller synthesized beam). Moreover, larger beams also imply larger foregrounds and larger confusion from background sources. Since it is not obvious \emph{a priori} whether single-dish telescopes or radio interferometers will perform better, we consider results for both observing modes. 
 
In the near future, one of the most powerful radio telescopes available will be the SKA. We consider here a total of five different configurations for SKA~\cite{WinNT}. Phase one of SKA-Mid (labeled here as `SKA1-Mid') 
will be built and operational as early as 2022. We also consider a configuration of SKA-Mid consistent with the proposed upgrade (labeled here as `SKA2-Mid'), which has a slightly wider frequency band, 10 times more telescopes, and includes phase array feed (PAF) technology.\footnote{A PAF consists of an array of receivers that are off-set in the focal plane of the dish and therefore see slightly different parts of the sky. Combining multiple simultaneous beams, an antenna equipped with PAF provides a much larger field-of-view.} The analysis below allows both telescopes to operate in either single-dish or interferometric modes. The SKA collaboration has also planned the construction of a low frequency array, which is assumed here to operate solely in the interferometric configuration (note that the synthesized beam increases at small frequencies, and thus the potential tradeoff between interferometric and single-dish observations is reduced significantly). In the coming years, the radio community will begin a significant experimental effort to map large scales at frequencies below 1 GHz in connection to the study of cosmological hydrogen (redshifted 21 cm line). Data collected in these surveys can also be used to search for the signal discussed here. As a reference telescope of this class, we consider the inteferometer HIRAX~\cite{Newburgh:2016mwi}. Other planned surveys that can be used to this aim involve, among others, the APERTIF, BINGO, CHIME, FAST and Tianlai telescopes (see \cite{Bull:2014rha} for a review of forthcoming experiments). 

Details of the performances for HIRAX and for the various configurations of SKA considered in this work are summarized in \Tab{tab:experiments}.

The axion signal is expected to appear as a narrow spectral line, broadened by the axion velocity dispersion. For targets in ordinary galaxies, the expected velocity dispersion is $\sim 10^{-3}\,c$, while for dwarf galaxies may be as small as a few km/s (i.e., a few times $10^{-5}\,c$). An experiment hoping to resolve the spectral features of the line would therefore typically require, respectively, $\sim 10^3$ and $\sim 10^5$ frequency channels. This is in full compliance with SKA capabilities, and while the current HIRAX design includes only 1024 channels, there exists however a foreseen possibility to up-channelizing data to a spectral resolution of 1.5 km/s.

\subsection{Expected Flux}\label{sec:flux}
The flux density, \ie the power per unit area per unit frequency, from the spontaneous and stimulated decay of an axion is given by
\begin{equation}\label{eq:flux_decay}
S_{decay} = \frac{\Gamma_a}{4\pi \Delta\nu}\int \, d\Omega \, d\ell \, \rho_a(\ell, \Omega) \, e^{-\tau(m_a, \ell, \Omega)} \, \left[ 1 + 2 f_\gamma(\ell, \Omega, m_a) \right] \, ,
\end{equation}
where $m_a$ is the axion mass, $\Gamma_a = \tau_a^{-1}$ is the spontaneous decay rate of axions (given by the inverse of \Eq{eq:tau}), $\rho_a(\ell, \Omega)$ is the axion mass density, $\Delta \nu$ is the width of the axion line, $f_\gamma(\ell, \Omega, m_a)$ is the ambient photon occupation evaluated at an energy $E_\gamma = m_a/2$, and $\tau$ is the optical depth. The integral in \Eq{eq:flux_decay} should be performed over the solid angle covered by the radio telescope and the line of sight between the source and the location of Earth. 

\Eq{eq:flux_decay} describes an isotropic emission. On the other hand, if the ambient radiation field is anisotropic, then the stimulated axion-decay emission will follow the direction of the ambient radiation (since the photon emitted from stimulated axion-decay is produced in the same quantum state as the ambient photon sourcing the stimulated emission).
As it will be clearer in the following, we either consider radiation field that are isotropic up to very good approximation (such as CMB and extragalactic background) or model the ambient photons at the source location by exactly taking only the measured continuum emission, namely, only the photons directed towards us. In other words, we consider photons with the right direction to induce a stimulated emission directed towards our location and therefore our estimate are not affected by possible anisotropies in the considered ambient fields.

It is conventional in radio astronomy to work with effective temperatures rather than flux densities. The observed antenna temperature in a single radio telescope is given by
\begin{equation}\label{eq:antennaT}
T_{\rm ant} = \frac{A_{\rm eff} \, \left< S \right>}{2 k_b} \, ,
\end{equation}
where $A_{\rm eff}$ is the effective area of the telescope, which we set to be $A_{\rm eff}=\eta\,A_{\rm coll}$ where $A_{\rm coll}$ is the physical collecting area of the telescope and $\eta$ is the efficiency (assumed to be 0.8 for SKA~\cite{WinNT} and 0.6 for HIRAX~\cite{Newburgh:2016mwi}), and $\left<S\right>$ is the bandwidth-averaged flux density. Throughout this work we take the bandwidth to be equal to that width of the axion line, \ie $\Delta B = \Delta \nu = \nu_a \, \sigma/c$, where $\nu_a$ is the central frequency of the line and $\sigma$ the velocity dispersion of the dark matter particles. 

\subsection{Telescope Sensitivities}
\label{sec:telescopes}

\begin{table}[]
 \small
  \centering
 \renewcommand\arraystretch{2.2}
\begin{tabular}{|c||c|c|c|c|}
\hline
 & SKA1-Mid & SKA2-Mid & SKA-Low & HIRAX \\ \hline \hline
Freq. [GHz] & 0.35-14 & 0.35-30 & 0.05-0.35 & 0.4-0.8 \\ \hline 
$N_{\rm PAF}$& 1 & 36 & 1 & 1  \\ \hline 
$N_{tele}$ & 200 & 2000 & 911 & 1024  \\ \hline 
$D$ [m] & 15 & 15 & 35 & 6  \\ \hline 
$\theta_{\rm sinth} $ [$^\prime$] &3.6-0.09 & 3.6-0.04 & 25.2-3.6 & 10-5  \\ \hline 
$T_{rcvr}$ [K] & 20 & 20 & 40 &  50 \\ \hline 
\end{tabular} 
\caption{Telescope performances and configurations considered in this work. For the case of SKA1-Mid and SKA2-Mid, we consider the array working both in interferometric and single-dish modes.
}
\label{tab:experiments}
\end{table}

The minimum (rms) observable temperature for a single telescope and one polarization is given by
\begin{equation}\label{eq:antennaNoise}
T_{\rm min} = \frac{T_{\rm sys}}{\sqrt{\Delta B \, t_{\rm obs}}} \, ,
\end{equation}
 where $t_{\rm obs}$ is the observation time (set to be equal to 100 hours throught this work), $\Delta B$ is the bandwidth, and the system temperature $T_{\rm sys}$ is given by $T_{\rm sys}=T_{\rm rcvr}+T_{\rm sky}$.
$T_{\rm rcvr}$ is the noise of the receiver, while $T_{\rm sky}(\ell,b)$ is the ``sky noise'' in the direction of observation.
In the case of dSph galaxies, we extracted the temperature $T_{\rm sky}$ at the dSph position from the Haslam map~\cite{Haslam:1982zz} at 408 MHz and rescaled to other frequencies with a spectral index of $-2.55$.
In the cases of the Galactic center and M87, we considered the temperature derived by the same observations we used to describe the radiation field in the context of computing the stimulated emission, see below. 
Finally, in the case of the Galactic halo, we adopt a sky-average value $T_{\rm sky}\simeq 60\,(\lambda/{\rm m})^{2.55}$ K \cite{WinNT}, since we are considering a very large fraction of the sky.

Transitioning from \Eq{eq:antennaT} and \Eq{eq:antennaNoise} to observable signal-to-noise of a telescope or an array depends inherently the mode of observation, \ie interferometric or single-dish observation.
The field of view (FoV) can be computed similarly in the two cases, while angular resolution and sensitivity have to be treated separately.

\subsection{Field of view}
The angle corresponding to the full-width at half maximum (FWHM) of the primary beam is given by
\begin{equation}\label{eq:angle_pix}
\theta_{pb} \simeq 1.22 \, \frac{\lambda}{D} \, \simeq 0.7^\circ \left( \frac{1 \, {\rm GHz}}{\nu}\right) \, \left(\frac{15 \, {\rm m}}{D} \right) \, ,
\end{equation}
where $\lambda$ and $\nu$ are the wavelength and frequency of observation, and $D$ is the diameter of the dish/station.
Here, we consider the primary beam area to be $\Omega_{pb}=2\pi\,(1- \cos(\theta_{pb}/2))$. In the cases of SKA1-Mid, SKA-Low and HIRAX, the FoV is set by the primary beam, \ie ${\rm FoV}=\Omega_{pb}$.

The FoV can be however enlarged by equipping the interferometer with PAF technology, which makes ${\rm FoV}=N_{PAF}\,\Omega_{pb}$, with $N_{PAF}$ expected to be $\gtrsim 36$ for next generation radio telescopes~\cite{WinNT}. This is the picture we consider for SKA2-Mid and in this case the signal-to-noise ratio becomes
\be
\left(\frac{S}{N} \right)^{PAF}=\sqrt{\sum_{i=1}^{N_{PAF}} \left(\frac{S}{N} \right)_{b_i}^2}
\label{eq:senspaf}
\ee
where $(S/N)_{b_i}$ is the signal-to-noise in the beam $i$ and there are $N_{PAF}$ beams in the FoV.
For a spatially uniform emission, the increase of the FoV due to PAF would lead to an increase of the signal-to-noise by a factor of ten.

\subsection{Single-dish angular resolution and sensitivity}\label{sec:single}
For single-dish telescopes, the angular resolution is set by \Eq{eq:angle_pix}.
The latter defines the integration angle to be used in \Eq{eq:flux_decay}, which leads (through Eq.~\ref{eq:antennaT}) to a certain $T_{\rm ant}^{\rm pb}$.
The signal-to-noise ratio for a single telescope and one polarization is simply given by the ratio between \Eq{eq:antennaT} and \Eq{eq:antennaNoise}:
\be
\left(\frac{S}{N} \right)_{sd,{\rm single}}=\frac{T_{\rm ant}^{\rm pb}}{T_{\rm min}}\;.
\label{eq:sens0}
\ee

Clearly the actual temperature measured by the telescope will not just include the emission associated to axion decay, but also a number of other Galactic or extragalactic radio sources.
We assume to be able to remove the continuum (``smooth'') radiation, since the telescope considered in this work have a large number of frequency channels that can be used to constrain the spectrum.
For a discussion about foreground removal, see, e.g., \cite{Santos:2015gra,Liu:2011ih}.
On top of that, if the continuum emission is also spatially smooth, like in the direction of dSphs (since they are not expected to source a significant continuum emission, see, e.g., \cite{Regis:2014koa}),
the foreground does not even enter interferometric observations which are blind to large scales. Therefore, the case of interferometric observations of dSph can be considered the most solid scenario for what concerns foreground removal.

The presence of spectral lines would instead constitute an irreducible background (in particular, if the width is comparable to the width of the axion line). However, there are only a very limited number of radio lines in the frequency range of interest. The only potentially problematic spectral line is the redshifted 21-cm line, which is relatively bright at low redshifts.
The observed frequency scales as $\nu_{obs}=\nu_{em}/(1+z)$ and the emission stays significant up to approximately $z\lesssim 5$. Therefore, it affects the possible detection of axions with masses between about 2-12 $\mu$eV. This range is however already strongly constrained by haloscopes except for a narrow window around 3-4 $\mu$eV (see Fig.~\ref{fig:resultsDW}). To study potential technical ways to remove such a background (exploiting \eg different line width and morphology of the axion signal) is beyond the goal of this paper.

Considering an array with $N_{\rm tele}$ telescopes observing in single-dish mode, the signal-to-noise is given by
\begin{equation}
\left(\frac{S}{N} \right)_{sd,\rm array} = \sqrt{N_{\rm tele} \, n_{\rm pol}}\,\left(\frac{S}{N} \right)_{\rm single}= \sqrt{N_{\rm tele} \, n_{\rm pol}}\,\frac{T_{\rm ant}^{\rm pb}}{T_{\rm min}}\, , 
\end{equation}
where $n_{\rm pol}$ is the number of polarizations (we set $n_{\rm pol}=2$).
Finally, for an array equipped with PAF the signal-to noise is simply
\begin{equation}
\left(\frac{S}{N} \right)_{sd,{\rm array}}^{PAF} = \sqrt{N_{\rm tele} \, n_{\rm pol}} \,\sqrt{\sum_{b=1}^{N_{PAF}}\left(\frac{T_{\rm ant}^{b}}{T_{\rm min}}\right)^2}\; , 
\end{equation}
where $T_{\rm ant}^{b}$ is the antenna temperature in the beam $b$.

\subsection{Angular resolution and sensitivity of interferometers}\label{sec:inter}
In a radio interferometer, the angular resolution is set by the longest baseline $b_{\rm max}$, \ie $\theta_{\rm res} \simeq 1.22 \,\lambda/b_{\rm max}$, while the largest scale that can be imaged is set by the shortest baseline $b_{\rm min}$, \ie $\theta_{\rm max} \simeq 1.22 \,\lambda/b_{\rm min}$.
The actual synthesized beam and largest observable scale depends also on observational details, such as the coverage in the visibility plane.
The emissions discussed here have a typical size $\lesssim \theta_{\rm max}$ for all telescopes, so it is reasonable to assume that there is no flux lost by the interferometric observations, except possibly for high frequencies and extended targets (cases for which we will consider single-dish observations, see~\Sec{sec:results}).
The synthesized beam is determined by considering an ``average'' baseline, leading to the values of $\theta_{\rm sinth}$ detailed in Table~\ref{tab:experiments}.
The signal in each ``pixel'' is the integral of \Eq{eq:flux_decay} over the synthesized beam, which provides $T_{\rm ant}^{\rm sb}$ (again through \Eq{eq:antennaT}).

Considering an array with $N_{\rm tele}$ telescopes operating in an interferometric mode, there are $N_{\rm tele} (N_{\rm tele} - 1)/2$ independent baselines, the signal-to-noise given by
\begin{equation}
\left(\frac{S}{N} \right)_{if,\rm array} = \sqrt{\frac{1}{2}N_{\rm tele} (N_{\rm tele} - 1) \,n_{\rm pol}} \, \sqrt{\sum_{pix=1}^{N_{pix}}\left(\frac{T_{\rm ant}^{pix}}{T_{\rm min}}\right)^2}\; ,
\label{eq:interf}
\end{equation}

where $N_{pix}$ is the number of synthesized beams contained in the area of the primary beam.
For a spatially uniform emission, the signal-to-noise is increased by a factor $\sqrt{N_{pix}}$ in the limit that the synthesized beam is equal to the primary beam.

To compute $\left(\frac{S}{N} \right)_{if,{\rm array}}^{PAF}$, one can again apply \Eq{eq:senspaf}.

\subsection{Radiation Fields for Stimulated Emission}\label{sec:radfields}

\begin{figure}[]
\begin{center}
\includegraphics[width= 0.48 \textwidth]{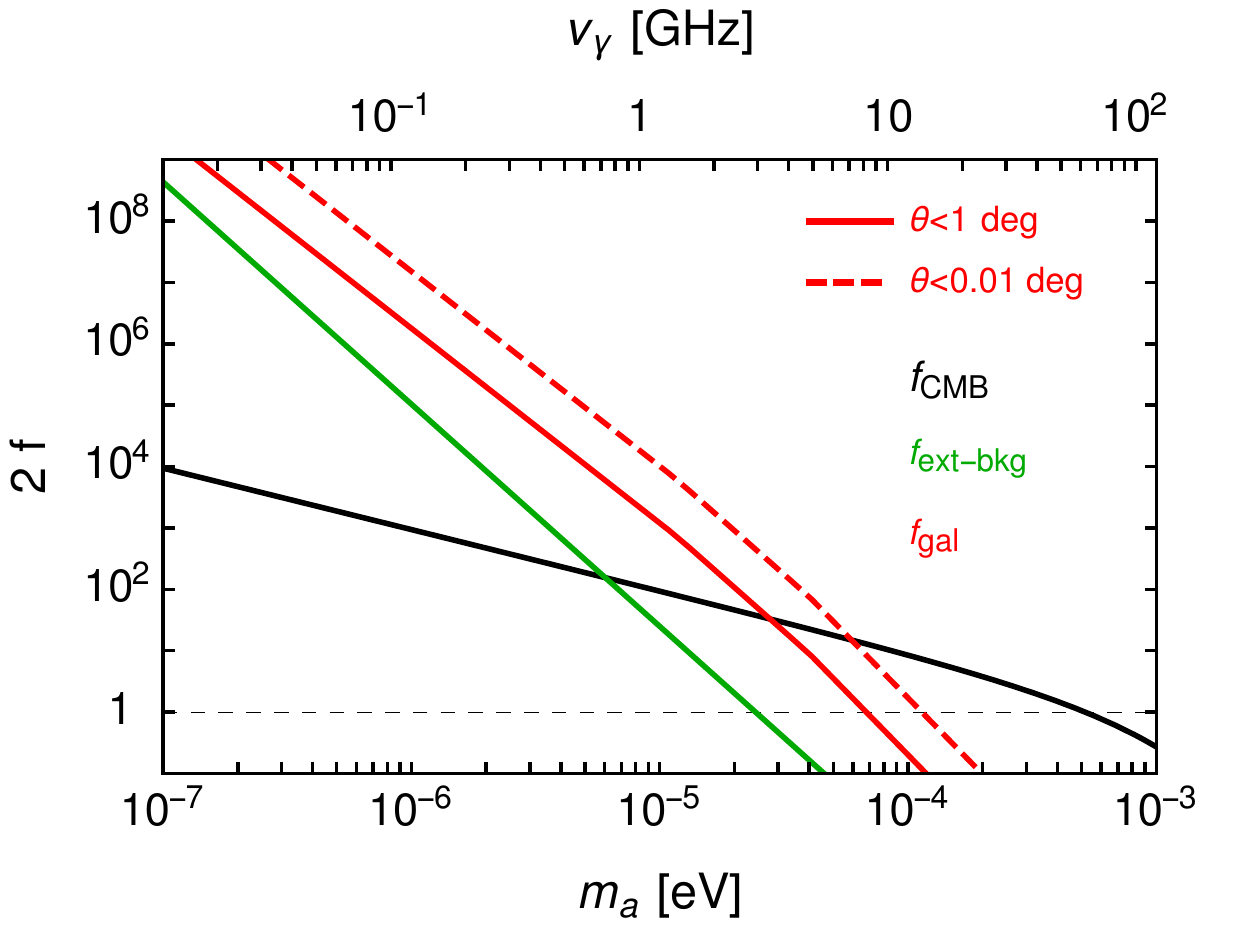}
\includegraphics[width= 0.464 \textwidth]{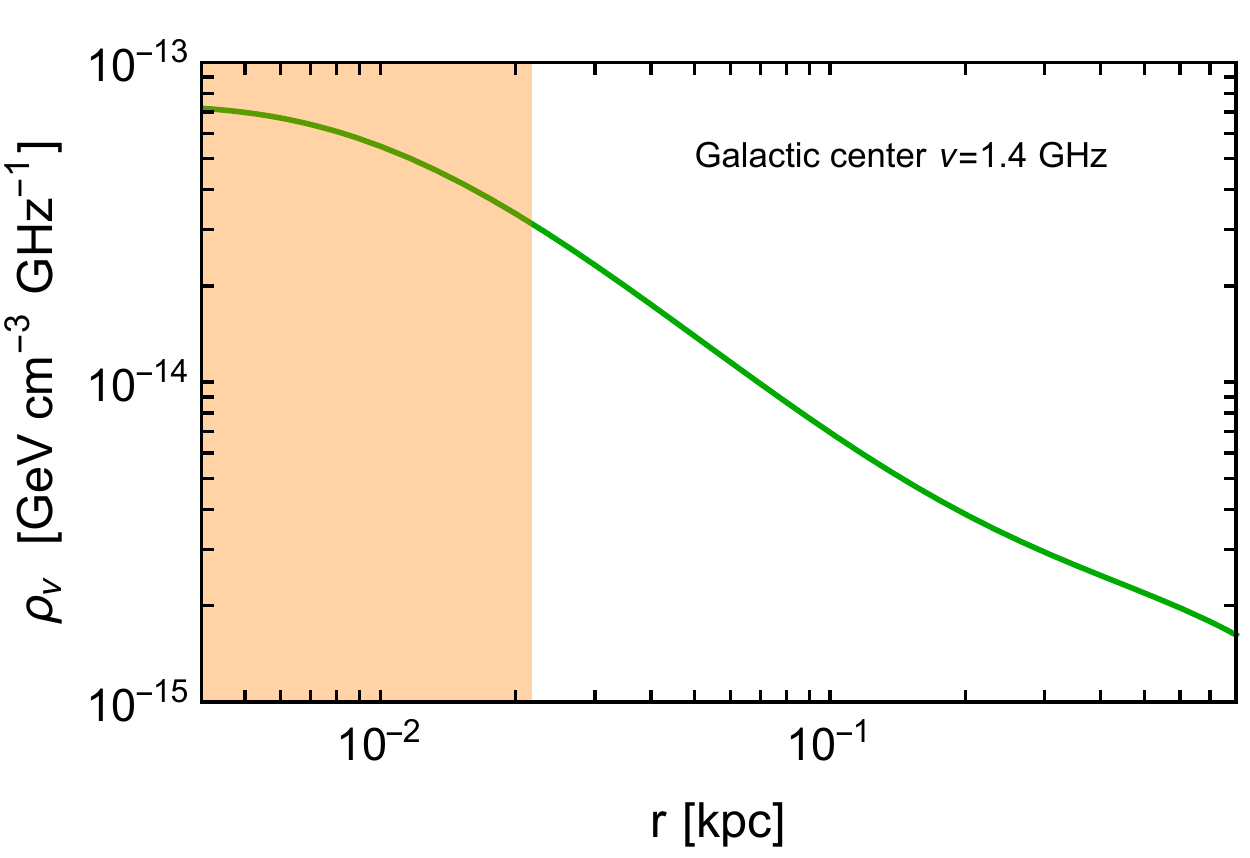}
\caption{Left: Stimulated emission factor ($2\, f_\gamma$)
broken down in terms of contributions from the CMB, the extragalactic radio background, and Galactic diffuse emission. The Galactic contribution is averaged in a region of angular radius of 1 and 0.01 degrees about the Galactic center, see~\Eq{eq:galStimulated}.
Right: Differential photon energy density at $\nu = 1.4$ GHz as a function of the distance from the Galactic Center. }
\label{fig:stimulated}
\end{center}
\end{figure}

As discussed in \Sec{sec:stimulated}, the presence of a non-negligible photon background with the same energy as that produced in the axion decay implies a stimulated enhancement of the axion decay rate; this effect manifests in terms of a non-zero contribution to $f_\gamma$ in \Eq{eq:flux_decay}. In general, $f_\gamma$ will be a linear combination over the sources which contribute to the photon bath. At radio frequencies, this includes, but is not limited to\footnote{It is worthwhile to note that at high-frequencies, free-free emission in hot and high-density environments may become the dominant contribution to $f_\gamma$. This contribution is neglected in this work. }, photons from the cosmic microwave background (CMB), diffuse emission from within the galaxy under consideration, and the radio background from extragalactic sources, 
\ie we assume
\begin{equation}\label{eq:fgamma_general}
f_\gamma(\ell, \Omega, m_a) \simeq f_{\gamma, {\rm CMB}}(m_a) + f_{\gamma, {\rm gal}}(\ell, \Omega, m_a) + f_{\gamma, {\rm ext-bkg}}(m_a) \, ,
\end{equation}
where the spatial dependence of $f_\gamma$ appears exclusively in the contribution from galactic emission, while $f_{\gamma, {\rm CMB}}$ and $f_{\gamma, {\rm ext-bkg}}$ are (at first approximation) isotropic.

The photon occupation from a blackbody spectrum is given by
\begin{equation}\label{eq:blackbody}
f_{\gamma, bb} = \frac{1}{e^{x} - 1} \, \hspace{1cm} x \equiv \frac{E_\gamma}{k_b T} 
\end{equation}
where $k_b$ is Boltzmann's constant and $T$ is the blackbody temperature. \Eq{eq:blackbody} can be used to incorporate the contribution to $f_\gamma$ from the CMB, taking $T = 2.725$ K. 

For the galactic diffuse and the extragalactic contributions we use \Eq{eq:ffromI}, namely, we derive the stimulated enhancement factor from the measured radio intensity.
The extragalactic radio background has been measured \cite{Fixsen:2009xn,Fornengo:2014mna} to have a frequency-dependent temperature given by
\begin{equation}
T_{\rm ext-bkg} (\nu) \simeq 1.19 \, \left(\frac{{\rm GHz}}{\nu} \right)^{2.62} \, {\rm K} \, .
\end{equation}
The contribution to $f_\gamma$ from galactic diffuse emission will be instead specified for each target (Galactic center, M87, Galactic halo) in the next Section. 
Its spatial dependence is computed from the angular profile of the measured radio flux. 

\section{Results}
\label{sec:results}

Below we present the projected 
sensitivity contours in the ALP parameter space for different astrophysical targets.
For the SKA-Mid, we considered the telescope operating both in single-dish and interferometric modes.  
These two configurations lead to similar results for the Galactic center and dwarf spheroidal galaxies and we quote only the sensitivities from single-dish observations.
This is because in the high-frequency end, the largest scale that can be imaged by the interferometer becomes comparable/smaller than the size of the source, so there might be some loss of flux (that we are not including in the modeling), which does not happen in the case of single-dish observations.
At lower frequencies, observational beams become larger, and SKA-Low and HIRAX do not face above issue.
For M87 instead, SKA-Mid operating in interferometric mode provides significantly better sensitivities, therefore we show the results only for this configuration. 
Given the distance from us and the large  stimulated emission factor in the central region of the galaxy, the emission from M87 is more compact than in the case of the Galactic center and dwarf spheroidal galaxies. This explains why the interferometric mode, which has a better rms sensitivity and a smaller smaller synthesized beam, is favored with respect to the single-dish mode.

\begin{figure}[]
\begin{center}
\includegraphics[width= 0.48 \textwidth]{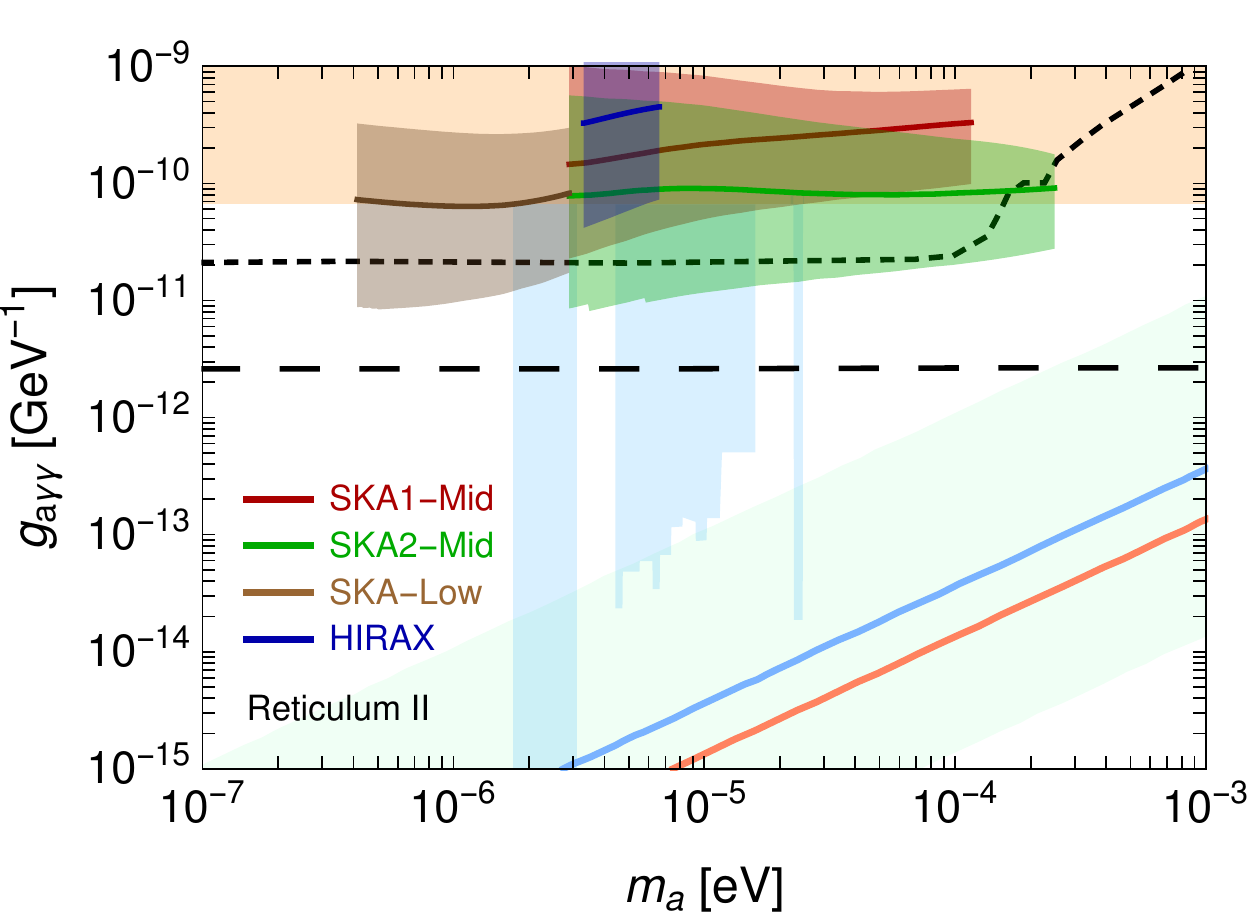}
\includegraphics[width= 0.48\textwidth]{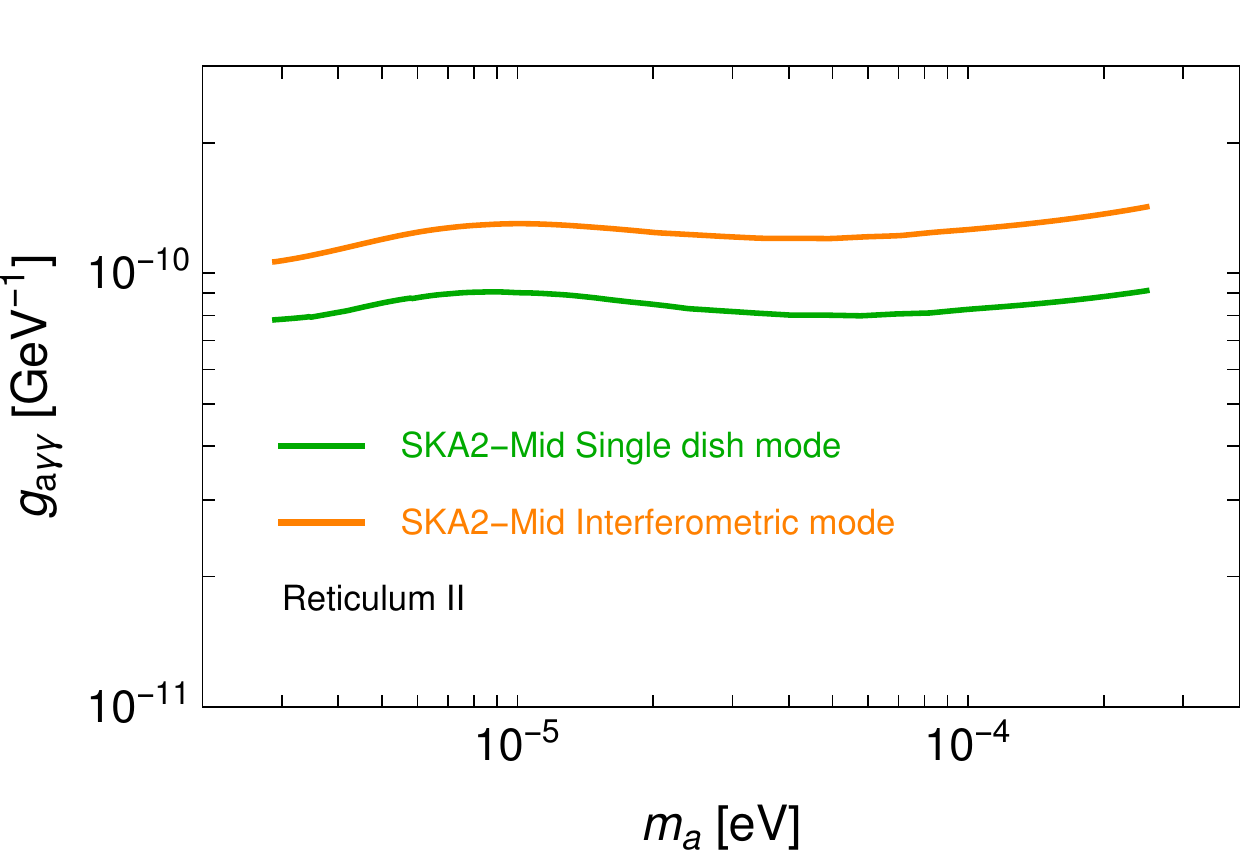}
\caption{Left panel: Projected sensitivities for the Reticulum II  dwarf galaxy. Results are displayed alongside current bounds from haloscopes (light blue)~\cite{Hagmann:1998cb,Asztalos:2001tf,Du:2018uak,Zhong:2018rsr} and helioscopes (orange)~\cite{Anastassopoulos:2017ftl}, projected bounds from ALPS-II~\cite{Bahre:2013ywa} (black, short dashed) and IAXO~\cite{Irastorza:2013dav,Armengaud:2014gea} (black, long dashed), and benchmark QCD axion models (light green band, blue line, orange line)~\cite{DiLuzio:2016sbl}. The width of the expected exclusion contours reflect the astrophysical uncertainty in the dSph environment. Right panel: Comparison between the sensitivites of interferometric and single-dish observations for the Reticulum II dwarf galaxy, considering the SKA2-Mid configuration. }
\label{fig:resultsDW}
\end{center}
\end{figure}

\subsection{Dwarf galaxies}
\label{sec:dwarf}

As a first target we consider dwarf spheroidal galaxies, refining the analysis performed in \cite{Caputo:2018ljp}. 
Dwarf spheroidal galaxies offer large dark matter densities, a low velocity dispersion and their angular size is within the field of view of the radio telescopes that we are considering.
These properties make them prime targets for searches of radio emissions produced by the decay of axions.
The signal, as shown in~\Eq{eq:flux_decay}, depends on the integral of the dark matter density distribution in the region of observation, the so-called  D-factor $D(\theta):$
\begin{equation}
	D(\theta) =\int d\Omega\, dl \, \rho_a(l,\Omega),
\end{equation}
with $\theta$ the angular distance from the center of the the dwarf galaxy.
A significant effort is ongoing to reconstruct the D-factors from stellar kinematical observations, see for example~\cite{Geringer-Sameth:2014yza, Bonnivard:2015tta, Bonnivard:2015xpq, Hayashi:2016kcy, Sanders:2016eie, Evans:2016xwx}. Here we make use of the publicly available data in~\cite{Bonnivard:2015xpq, Bonnivard:2015tta}, which provide the median values of the D-factors and their uncertainty intervals for several dwarf galaxies.
In particular we focus on Reticulum II, which is one of the most promising dwarf galaxies that can be observed by radio telescopes located in the southern hemisphere, like SKA and HIRAX.
Similar D-factors have been reported also for other observable (by southern located telescopes) dwarf spheroidal galaxies, both classical (e.g. Sculpture) and ultrafaint (e.g. Coma).

The dark matter velocity dispersion is estimated from the measured stellar velocity dispersion which we take to be $\sigma  \simeq 4$ km s$^{-1},$ see\cite{Bonnivard:2015xpq}.
The uncertainty on $\sigma$ affects the results mildly, since the sensitivity on $g_{a\gamma\gamma}$ scales as $\sigma^{-1/4}.$
For the stimulated emission, we include only the contributions from the CMB and the extragalactic radiation, neglecting any (currently undetected, and likely subdominant) radio emission produced inside the dwarf galaxy.

We show our results in the left panel of~\Fig{fig:resultsDW}. The bands refer to the 95\% credibility interval on the D-factor provided in~\cite{Bonnivard:2015xpq}.
Under the stated assumptions future radio telescopes might be able to probe a currently unexplored region of the parameter space, corresponding to $g_{a\gamma\gamma}\gtrsim 10^{-11}$ GeV$^{-1}$.

In~\Fig{fig:resultsDW} (right), we compare the results of single-dish versus inteferometric observing modes for the SKA2-Mid configuration.
As clear from the plot, the difference is limited. For what said in \Sec{sec:single}, this also shows that the main conclusion of this paper should be not affected by the removal of the continuum emission.

The sensitivity reported in this work differs by about three orders of magnitude (about a factor of 30 in the axion-photon coupling) with respect to \cite{Caputo:2018ljp}. That work applied the SKA interferometric sensitivity as if the source were point-like rather than extended as instead we consider in~\Eq{eq:interf}. Moreover, they considered an optimistic SKA collecting area which is not currently foreseen in the future SKA design (i.e., $A_{eff}/T_{sys}=10^5 {\rm m^2/K}$ at high-frequency, about one order of magnitude larger than our SKA-2 case).


\subsection{Galactic center}
\label{sec:gc}

Due to the close proximity and high dark matter column density, the Galactic center is typically among the most promising targets for indirect dark matter searches. Additionally,  the presence of a large synchrotron background is expected to lead to a significant enhancement of the decay rate of axions. We describe the details and caveats of this calculation below.

The distribution of dark matter in the Galactic center, and in particular the inner slope of the density profile, remains largely unknown. N-body simulations of collisionless cold dark matter predict density profiles well-modeled by a Navarro-Frenk-White (NFW) profile~\cite{Navarro:1995iw}, in which $\rho(r) \propto r^{-1}$ at small radii. 
However, mechanisms have been proposed which can either flatten or steepen the distribution to produce a core or cusp. In an attempt to account for this source of uncertainty, we present sensitivity studies for three distinct profiles,  one `reference' distribution, and two profiles intended to characterize the relative extremes. The reference distribution that we consider is the NFW profile:
\begin{equation}\label{eq:nfw}
\rho(r) = \frac{\rho_s}{\left(\frac{r}{r_s}\right)\left(1 + \frac{r}{r_s}\right)^{2}} \, ,
\end{equation}
where we take a scale radius $r_s=24.42$ kpc~\cite{Cirelli:2010xx} and $\rho_s$ is normalized in order to obtain a density at the Earth's location ($r_{\odot}=8.3$ kpc) $\rho(r_{\odot})= 0.3 \,{\rm GeV / cm^3}.$ 
For the more optimistic scenario we consider a cuspy distribution, given by a generalized NFW profile
\begin{equation}\label{eq:gen_nfw}
\rho(r) = \frac{\rho_s}{\left(\frac{r}{r_s}\right)^{\gamma}\left(1 + \frac{r}{r_s}\right)^{3 - \gamma}} \, ,
\end{equation}
with $r_s$ and $\rho_s$ defined as before, and an inner slope $\gamma$ taken to be 1.3. 
To model the cored dark matter density profile we adopt the so-called Burkert profile, given by
\begin{equation}
\rho(r) = \frac{\rho_s}{\left(1 + \frac{r}{r_{sb}}\right)\left(1 + \frac{r}{r_{sb}}\right)^2} \, ,
\end{equation}
where $r_{sb} = 12.67$ kpc~\cite{Cirelli:2010xx}, and as before the scale density is set to provide the reference local density $\rho(r_{\odot}).$ 

The velocity dispersion of dark matter, characterizing the width of the decay line, is taken to be $\sigma=200$ km/s. 
To estimate the Galactic contribution to $f_\gamma$ in~\Eq{eq:fgamma_general} the morphology of the radio diffuse emission should be taken into account.
Here, we attempt to infer the value and spatial dependence of $f_{\gamma, {\rm gal}}$ from the measurements of the radio flux in the Galactic center region presented
in~\cite{YusefZadeh:2012nh}. 
Specifically, we analyze the radial profile at $\nu_*=1.415$ GHz derived by averaging the emission in elliptical annuli (with aspect ratio of two-to-one), located around the Galactic center.
To simplify the analysis, we assume that the flux observed in~\cite{YusefZadeh:2012nh}, extending up to an angular scale of one degree along the semi-major axis, can be mapped onto a spherically symmetric region, \ie a circle rather than elliptic annuli, of equivalent area. The observed flux at Earth from a spherically symmetric emissivity $j(r)$ is given by
\begin{equation}
I(\theta) = \frac{1}{4\pi} \int \, ds \, j(\hat{r}(\theta, s, r_\odot)) \, ,
\end{equation}
where the line-of-sight coordinate $s$ is related to the radial distance from the Galactic center $\hat{r} = \sqrt{s^2 + r_\odot^2 - 2 \,s \,r_\odot \cos(\theta)}$ and $\theta$ is the aperture angle from the line-of-sight and the Galactic center direction. For the small angles $\theta$ under consideration, one can perform an Abel transform to infer the value of $j(r)$ from the value of $I(\theta)$ provided in~\cite{YusefZadeh:2012nh}.
Then, once the emissivity profile has been obtained, one can compute the differential photon density (at the frequency $\nu_*$) as a function of the distance $r$ from the Galactic center:
\begin{equation}
\rho_{\nu}(r) = \frac{1}{4\pi} \int \, ds \, d\Omega \, j(\hat{r}(\theta, s, r)) \, .
\label{eq:fgal}
\end{equation}
The resulting distribution is shown in the right panel of~\Fig{fig:stimulated}. The intensity profile $I(\theta)$ in~\cite{YusefZadeh:2012nh} flattens at small values of $\theta,$ an effect due to the finite angular resolution ($539^{\prime\prime}$) of the observations. In~\Fig{fig:stimulated}, the corresponding range of Galactic distances $r$ is shaded: in that region the $\rho_{\nu}(r)$ distribution is likely to be steeper than reported, but we conservatively choose not to extrapolate it.
The $\rho_{\nu}(r)$ is instead extrapolated as $r^{-2}$ at large distances, beyond the range covered by the observations. We checked that our results do not depend on the specific prescription adopted.
Finally, it is then straightforward to obtain the photon occupation number  $f_{\gamma, {\rm gal}}(r,\nu)$ from the density distribution $\rho_{\nu}(r)$.
The frequency dependence of the occupation number can be obtained from the observed spectral shape of the emission. From Table 1 of~\cite{YusefZadeh:2012nh} we obtain:
\begin{align}\label{eq:galsyn}
 f_{\gamma, {\rm gal}}(r,\nu) =  f_{\gamma, {\rm gal}}(r)\Big|_{\nu = \nu_*} \times \left\{ \begin{array}{cc}  \vspace{.2cm}
                (\nu / \nu_*)^{-3.173} & \hspace{5mm} \nu < \nu_*  \\ \vspace{.2cm}
                (\nu / \nu_*)^{-3.582} & \hspace{5mm}  \nu_*\leq  \nu \leq 4.85 \,{\rm GHz} \\ \vspace{.2cm}
               0.49 \times (\nu /\nu_*)^{-4.14} & \hspace{5mm} \nu > 4.85  \,{\rm GHz}\\
                \end{array} \right. \, .
\end{align}
The contribution of  $f_{\gamma, {\rm gal}}$ to the stimulated emission is presented in the left panel of~\Fig{fig:stimulated}. We show the average value in a region of $1$ and $0.01$ degrees around the Galactic center, defined as:
\begin{align}\label{eq:galStimulated}
\bar{f}_{\gamma, {\rm gal}} = \frac{\int \, d\Omega \, d\ell \, \rho_a(\ell, \Omega) \, f_\gamma(\ell, \Omega, m_a)}{\int \, d\Omega \, d\ell \, \rho_a(\ell, \Omega)}
\end{align}
The Galactic contribution dominates over the CMB and extragalactic ones in a large range of frequencies.

The free-free self absorption becomes relevant only at frequencies $\nu \lesssim10-20$ MHz for observations of targets located above the Galactic plane, as dwarf galaxies, and so can be neglected for our purposes.
Instead, the emission of sources located at low Galactic latitudes is significantly absorbed already at $\nu \lesssim200$ MHz, because of the large column density of electrons lying in the Galactic plane. We need to incorporate this effect in our analysis of the Galactic center. We compute the optical depth $\tau$ in~\Eq{eq:flux_decay} using~\cite{1974ApJ194715G} and setting a kinetic temperature of $5000$ K and an emission measure $EM=10^4$ cm$^{-6}$ pc.
The impact of absorption in our results, presented in the left panel of~\Fig{fig:resultsGCM87}, can be easily recognized: the sensitivity quickly degrades moving towards low frequencies. 
Overall, the sensitivity reach that we obtain is similar to the one found in~\Sec{sec:dwarf} for dwarf spheroidal galaxies.

\subsection{M87}
\label{sec:M87}

\begin{figure}[]
\begin{center}
\includegraphics[width= 0.48 \textwidth]{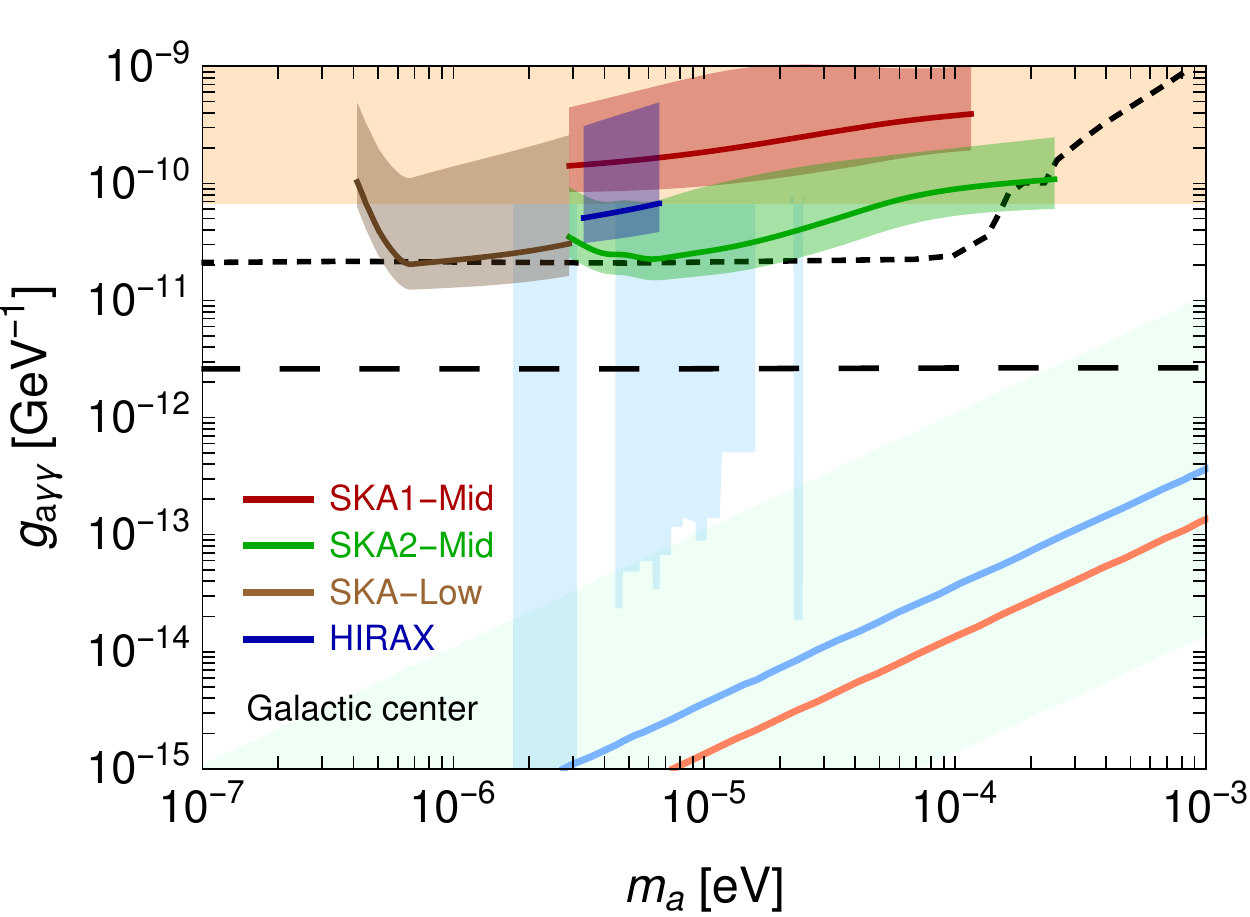}
\includegraphics[width= 0.48 \textwidth]{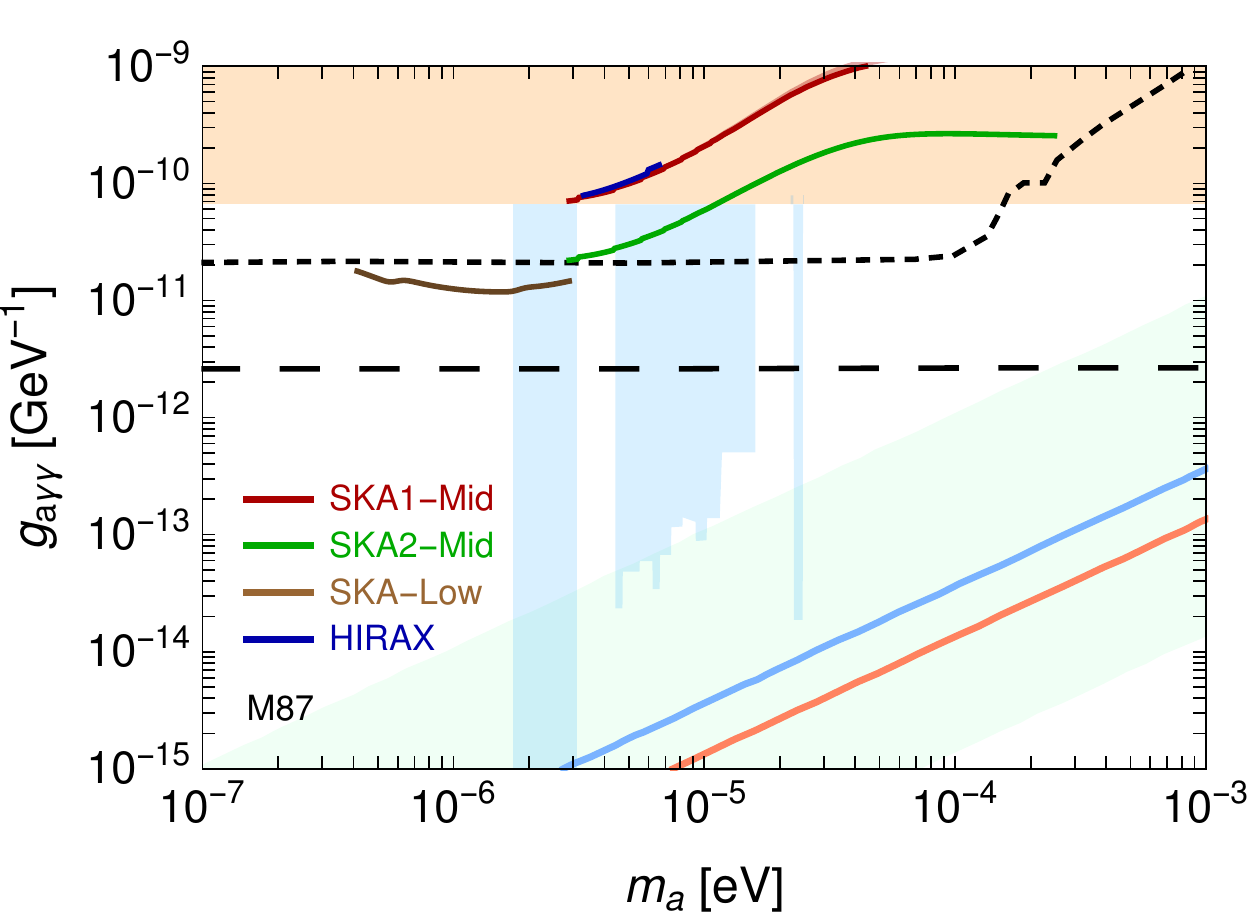}
\caption{Same as \Fig{fig:resultsDW} but for Galactic center (left) and M87 (right).}
\label{fig:resultsGCM87}
\end{center}
\end{figure}

Given its position in the sky, mass, and distance, the Virgo cluster is likely to be among the most promising galaxy clusters in the search for axion decay.
The massive elliptical galaxy M87 lying at its center accounts for a significant fraction of the Virgo mass ($\simeq 5-10$\%), and it is a bright radio emitter. 
An intra-cluster large scale radio halo is much fainter~\cite{Vollmer:2004hn}. 
We checked that M87 provides stronger constraints on ALP parameter space than the Virgo intra-cluster medium.
This is because the former hosts a large density of radio photons, leading to a large stimulated emission. In the following we show results only for M87.

To obtain the contribution to the stimulated emission from the diffuse radiation in M87, $f_{\gamma, {\rm gal}}(r,\nu)$ in~\Eq{eq:fgamma_general}, we proceed as follows. The total power radiated by M87 from 10 MHz up to 150 GHz is $L_{\gamma}=9.6\times10^{41}$erg/s, and it is predominantly produced within $r_p=40$ kpc of the center of the galaxy~\cite{Owen:2000vi}.
The photon energy density $\rho$ can be estimated simply as $\rho=\frac{L_{\gamma}\,r_p}{V\,c}$ with $V=\frac{4}{3}\pi r_p^3.$
Then, using a spectral dependence of the flux $\propto \nu^{-1}$ ~\cite{Owen:2000vi}, one can infer the photon energy density, and thus the occupation number $f_{\gamma, {\rm gal}}(r,\nu)$, at any frequency. For simplicity we consider a constant photon distribution  inside $r_p,$ with an average value estimated as explained above, and an abrupt depletion of the photon density outside this region.

The dark matter density distribution is modeled using the results of~\cite{oldham2016galaxy}, where the mass density has been inferred by jointly analyzing the dynamics of stars, globular clusters, and satellites. In particular, we consider two density distributions, the NFW profile and a cored profile (cgNW in~\cite{oldham2016galaxy}).
Finally we caution that we do not attempt to model any absorption inside M87. On the other hand, this can have an impact on both the derived energy density of background photons and the estimated axion-induced flux, with the two effects expected to be of similar size.

The sensitivities are shown in~\Fig{fig:resultsGCM87} (right). The changes due to the different choice of dark matter density profile are modest. The two  adopted profiles differ little at the distances corresponding to the angular scales under examination.

\subsection{Wide field surveys}
\label{sec:otherstar}

One of the primary goals of HIRAX is to observe the large-scale structure of the Universe through the 21 cm emission line produced by the neutral hydrogen, the so-called hydrogen intensity mapping. For this purpose HIRAX has been designed to measure a large fraction of the sky ($1.5\times 10^4$ squared degrees) with a fairly large integrated time ($10^4$ hours)~\cite{Newburgh:2016mwi}. We investigate whether the same observational campaign can be used to search for radio lines produced by the decay of axions inside the Galactic halo.
We estimate the sensitivity of HIRAX, approximating the region of observation with a circle centered at the Galactic center and spanning an equal area (i.e. $1.5\times 10^4$ squared degrees). The signal is computed assuming the NFW density profile in~\Sec{sec:gc} and a velocity dispersion $\sigma\simeq200$ km/s. To model the stimulated emission, in~\Eq{eq:fgamma_general} we incorporate the sum of the Galactic and extragalactic contribution using $T_{\rm sky}$ described in~\Sec{sec:telescopes} as a measure of the total radio flux. 

We find that the reach is at the level of the region already excluded by haloscopes, namely $g_{a\gamma\gamma}\simeq 8.6-17\times10^{-11} \, {\rm GeV}^{-1}$ 
in the mass range $m_a\simeq3.3-6.7\,\mu{\rm eV}.$ We have obtained similar sensitivities for the CHIME telescope~\cite{Newburgh:2014toa}, which will perform a complementary survey of the northern sky.

These estimates, although involving several approximations (namely the patch of the sky which will actually be observed, and the modeling of $f_{\gamma, {\rm gal}}(r,\nu)$), are enough to conclude that the strategies considered in the previous Sections, i.e. observations of the Galactic center, dwarf galaxies and galaxy clusters, are more promising to look for the decay of axions.  

The cumulative line emission from all dark matter halos at all redshifts forms a nearly isotropic emission with a continuum spectrum. We find the cosmological emission to be $\lesssim 10^{-4}$ of the measured extragalactic background (for couplings in the allowed range). Since this collection of lines determines a contribution with no prominent spatial or spectral features, it can be very complicated to identify. A potentially interesting way to overcome this issue is given by line-intensity mapping~\cite{Creque-Sarbinowski:2018ebl}. We postpose a dedicated analysis of this approach to future work.  

\subsection{Discussion}
\label{sec:othersmod}

The main goal of this work was to study the observability of the two-photon decay of ALPs, with the QCD axion used as a very well-motivated benchmark.
More generally, our results apply to any scenario in which a light dark matter candidate (with mass in the range $0.1-100\, \mu{\rm eV}$) has a monochromatic decay to one or two photons. For instance, one could extend the ALP model that we have analyzed, supplementing the Lagrangian by a term $a \, F^{\mu\nu}\tilde{F}_{\mu\nu}^\prime$, where $\tilde{F}_{\mu\nu}^\prime$ is the dual of a new field strength tensor arising from a dark $U(1)$ gauge symmetry\cite{Ilten:2018crw, Lees:2014xha, Bauer:2018onh, Endo:2012hp, Kozaczuk:2017per}. Here, if the mass of the dark photon is less than the mass of the axion, the axion decay will proceed via $a \rightarrow \gamma \gamma^\prime$ (this model has beed studied \eg in \cite{Kaneta:2016wvf,Kohri:2017oqn,Daido:2018dmu,Kalashev:2018bra}).

\begin{figure}
\begin{center}
\includegraphics[width= 0.48\textwidth]{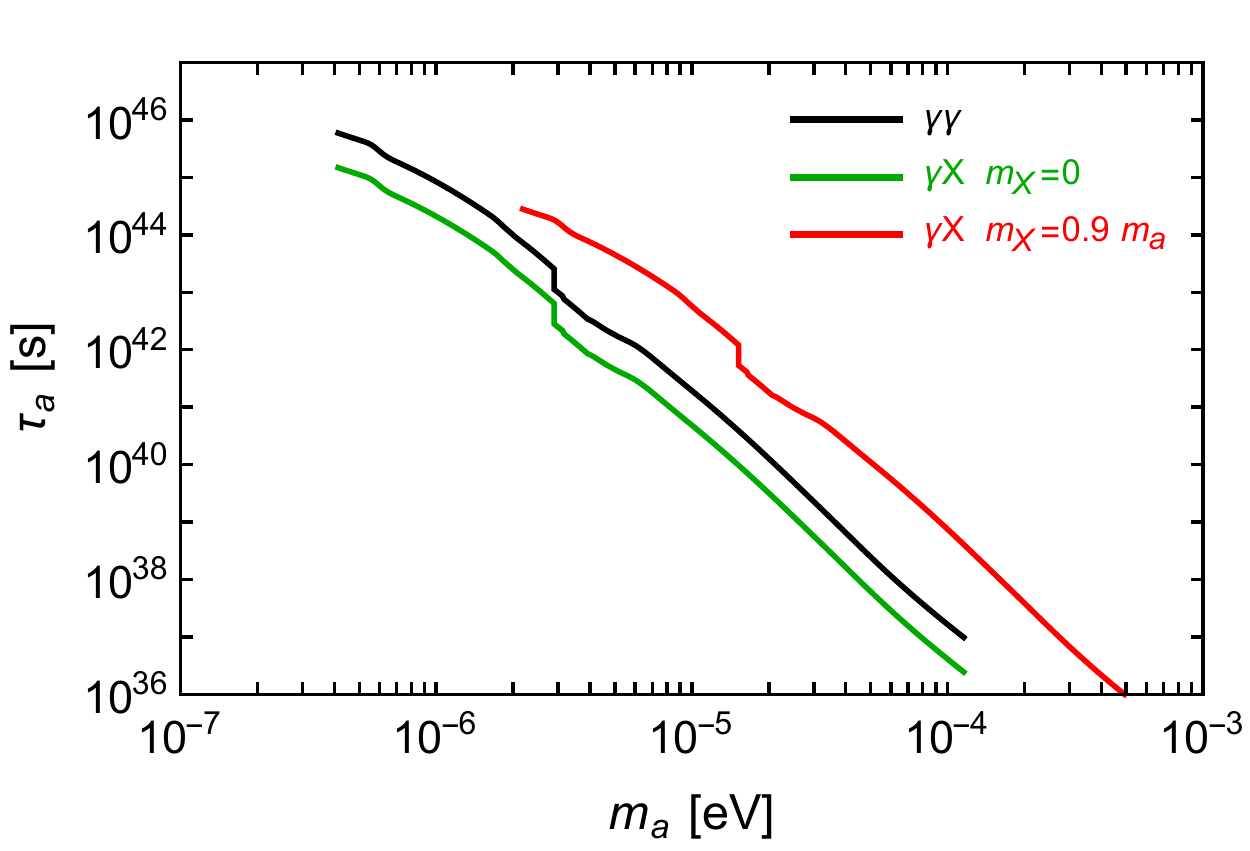}
\caption{\label{fig:decay} Projected sensitivity  
from Figs.~\ref{fig:resultsDW} and~\ref{fig:resultsGCM87}, translated into constraints on the ALP lifetime as a function of the ALP  mass. We show the cases of decays into two photons (black) and into a photon plus a state $X,$ massless (green) or with $m_X=0.9\,m_a$ (red).
}
\end{center}
\end{figure}

For sake of generality, in~\Fig{fig:decay} we present our sensitivities in terms of the ALP lifetime; these results can then be easily recast for alternatively models. 
The sensitivity curve is plotted as a function of the ALP mass. For the decay into two photons, we show the best sensitivity from Figs~\ref{fig:resultsDW} and~ \ref{fig:resultsGCM87}.  For the decay into a photon plus a lighter state $X,$ we recast the sensitivity using the relation $E_\gamma=\frac{m_a}{2}\left(1-\frac{m_X^2}{m_a^2}\right)$ and we show two cases: $m_X=0$ and $m_X=0.9\,m_a.$
Remarkably, \Fig{fig:decay} shows that it may be possible to probe lifetimes as large as $10^{46}$ seconds.

\section{Conclusions}
\label{sec:conclusions}

In this work we studied the radio emission arising from axion decays in various types of nearby astrophysical structures. We have presented projected sensitivities for targets with the best observational prospects, including the Galactic center, the ReticulumII dwarf galaxy, M87, and the Galactic halo. 
We have found that for ALPs with masses below meV, the stimulated decay arising from the presence of ambient photons results in a large enhancement of the decay rate -- potentially up to eight orders of magnitude for axion masses $\sim \mu$eV and in environments with large radio emission like the Galactic Center.

Once the axion mass and the coupling to photons are fixed, the main uncertainty comes from the dark matter distribution in the structure under consideration.
Indeed, the effect of stimulated emission can be well determined since the distribution of ambient photons can be confidently derived from continuum radio measurements. This is different from the possible signal coming from photon-axion conversion in strong magnetic fields around stars, which is potentially more promising but suffers of larger uncertainties associated with the poorly known astrophysics (namely the structure of the magnetic field and the plasma density).

We have showed that with near-future radio observations by SKA, it will be possible to increase sensitivity to the ALP-photon coupling by nearly one order of magnitude.
Interestingly, it has been shown that in this range of parameter space, axions provide a viable solution to a non-standard cooling mechanism identified in various stellar systems~\cite{Giannotti:2015kwo}.  
If forthcoming axion search experiments, such as ALPS-II and IAXO, find a signal consistent with axion dark matter in the $10^{-7}-10^{-3}$ eV mass range, the technique proposed here might become the standard route to understand the properties of dark matter, such as \eg its spatial distribution and clustering in cosmological structures.

\subsubsection*{Acknowledgments}

We thank S. Camera, J. Fonseca, N. Fornengo, P. Serpico and M. Viel.
MR acknowledges support by ``Deciphering the high-energy sky via cross correlation'' funded by Accordo Attuativo ASI-INAF n. 2017-14-H.0 and by the ``Departments of Excellence 2018 - 2022'' Grant awarded by the Italian Ministry of Education, University and Research (MIUR) (L. 232/2016). 
MR and MT acknowledge support from the project ``Theoretical Astroparticle Physics (TAsP)'' funded by the INFN. AC and SW acknowledge support from the European projects H2020-MSCAITN-2015//674896-ELUSIVES
and H2020-MSCA-RISE2015.

\footnotesize

\bibliographystyle{hunsrt}
\bibliography{biblio}
\end{document}